\documentclass[runningheads]{llncs}
\usepackage{graphicx}
\usepackage{color}
\usepackage{cite}
\usepackage{amsfonts,amssymb,amsmath}
\usepackage{multirow}
\usepackage{epsf}
\usepackage{longtable}
\usepackage{booktabs}
\usepackage[table]{xcolor}
\usepackage{ragged2e}
\usepackage{subfigure}
\usepackage{marvosym}
\usepackage{rotating}
\usepackage{hyperref}
\hypersetup{
colorlinks=true,
linkcolor=cyan,
filecolor=blue,      
urlcolor=red,	
citecolor=green,
}
\begin{document}
\title{TBraTS: Trusted Brain Tumor Segmentation}
%
\author{Ke Zou\inst{1,2}, Xuedong Yuan\inst{2}\textsuperscript{\Letter}, Xiaojing Shen\inst{3}, Meng Wang\inst{4}, Huazhu Fu\inst{4}\textsuperscript{\Letter}}
\authorrunning{K. Zou et al.}
\institute{National Key Laboratory of Fundamental Science on Synthetic Vision, Sichuan
University, Sichuan, China \and College of Computer Science, Sichuan University, Sichuan, China \and College of Mathematics, Sichuan University, Sichuan, China \and Institute of High Performance Computing, A*STAR, Singapore\\
\email{yxd@scu.edu.cn; hzfu@ieee.org}}

%
\maketitle              
\begin{abstract}
Despite recent improvements in the accuracy of brain tumor segmentation, the results still exhibit low levels of confidence and robustness. Uncertainty estimation is one effective way to change this situation, as it provides a measure of confidence in the segmentation results. 
In this paper, we propose a trusted brain tumor segmentation network which can generate robust segmentation results and reliable uncertainty estimations without excessive computational burden and modification of the backbone network. In our method, uncertainty is modeled explicitly using subjective logic theory, which treats the predictions of backbone neural network as subjective opinions by parameterizing the class probabilities of the segmentation as a Dirichlet distribution. Meanwhile, the trusted segmentation framework learns the function that gathers reliable evidence from the feature leading to the final segmentation results. Overall, our unified trusted segmentation framework endows the model with reliability and robustness to out-of-distribution samples. To evaluate the effectiveness of our model in robustness and reliability, qualitative and quantitative experiments are conducted on the BraTS 2019 dataset.

\end{abstract}
\section{Introduction}
Brain tumor is one of the most common brain diseases and can be classified as primary, brain-derived, and brain metastatic tumors. Among the primary malignancies, Gliomas with different levels of aggressiveness, accounting for 81\% of brain tumors \cite{14brainreview}. Multiple Magnetic Resonance Imaging (MRI) modalities that provide complementary biological information are one of the clinical tools for sensing tumor-induced tissue changes. Accurate segmentation of lesion areas from different imaging modalities is essential to assess the actual effectiveness before and after treatment.

Recently, many researchers have made great efforts to accurately segment the brain tumor from multimodal MRI images. Most of the methods are based on Convolutional Neural Networks (CNN)~\cite{chen2019disentangle,20attentionResBra, TransBTS}. U-Net~\cite{Unet15}  with skip-connections and its variants~\cite{AttentionU18,Unetadd20} are employed to improve performance of the brain tumor segmentation~\cite{dong17BraTS, 20attentionResBra}. Recently, highly expressive Transformer has been applied to medical image segmentation ~\cite{ji2022vps}, especially for brain tumor segmentation~\cite{TransBTS}. Although these models can improve the performance of segmentation, they are often prone to unreliable predictions. This is because these models use softmax to output predictions, which often leads to over-confidence, especially for error predictions~\cite{ICMLdeterministic20,2020trusted,2021trusted}. Moreover, clinicians often not only need accurate segmentation result, but also want to know how reliable the result is, or whether the specific value of the reliability of the result is high or low. Above observations have inspired us to develop models that can accurately segment tumor regions while providing uncertainty estimations for the segmentation results.

Uncertainty quantification methods mainly include dropout-based~\cite{14dropout, 16dropout}, ensemble-based~\cite{ensemble17}, evidential deep learning~\cite{evidential18}, and deterministic-based~\cite{ICMLdeterministic20}. At the same time, many works are devoted to associating the uncertainty with brain tumor segmentation. A simple way to produce uncertainty for brain tumor segmentation is to learn an ensemble of deep networks~\cite{2019ucBra, TMI20ensembleSeg}. On the downside, the ensemble-based methods require training the multiple models from scratch, which is computationally expensive for complex models.  Some brain tumor segmentation methods introduce the dropout in the test phase to estimate lesion-level uncertainties~\cite{2018effectDropSeg,MIA2020exploringDropSeg}. Despite this strategy reduces the computational burden, it produces inconsistent outputs~\cite{2018probabilisticU}. In addition, there is also a work that extends deep deterministic uncertainty~\cite{2021DetermSeg} to semantic segmentation using feature space densities. Although the above methods quantify the uncertainty of voxel segmentation, they all focus on taking uncertainty as feature input to improve segmentation performance rather than obtaining the more robust and plausible uncertainty. 

In this paper, we propose a \textbf{T}rusted \textbf{Bra}in \textbf{T}umor \textbf{S}egmentation  (TBraTS) network, which aims to provide robust segmentation results and reliable voxel-wise uncertainty for brain tumor. Instead of directly outputting segmentation results, our model enables the output of the underlying network in an evidence-level manner. This not only estimates stable and reasonable voxel-wise uncertainty, but also improves the reliability and robustness of segmentation. We derive probabilities and uncertainties for different class segmentation problems via Subjective Logic (SL)~\cite{SL16}, where the Dirichlet distribution parameterizes the distribution of probabilities for different classes of the segmentation results. In summary, our contributions are as follows: \\
(1) We propose a end-to-end trusted medical image segmentation model, TBraTS, for brain tumor aiming to quantify the voxel-wise uncertainty, which introduces the confidence level for the image segmentation in disease diagnosis.\\
(2) Our method could accurately estimate the uncertainty of each segmented pixel, to improve the reliability and robustness of segmentation.\\
(3) We conduct sufficient experiments on the BraTS2019 challenge to verify the segmentation accuracy of proposed model and the robustness of uncertainty quantification.\footnote{Our code has been released in \url{https://github.com/Cocofeat/TBraTS}. }

\section{Method}\label{S_2}
In this section, we introduce an evidence based medical image segmentation method which provides uncertainty for disease diagnosis. For the multi-modal voxel input, a backbone network is adopted to obtain segmentation results. Then, we elaborate on evidential deep learning to quantify the segmentation uncertainty to avoid high confidence values resulting from using softmax for prediction. At last, we construct the overall loss function.

\subsection{Uncertainty \& the theory of evidence for medical image segmentation} 
\begin{figure}[!t]
\centering
\includegraphics[height=1.5in,width=4.5in]{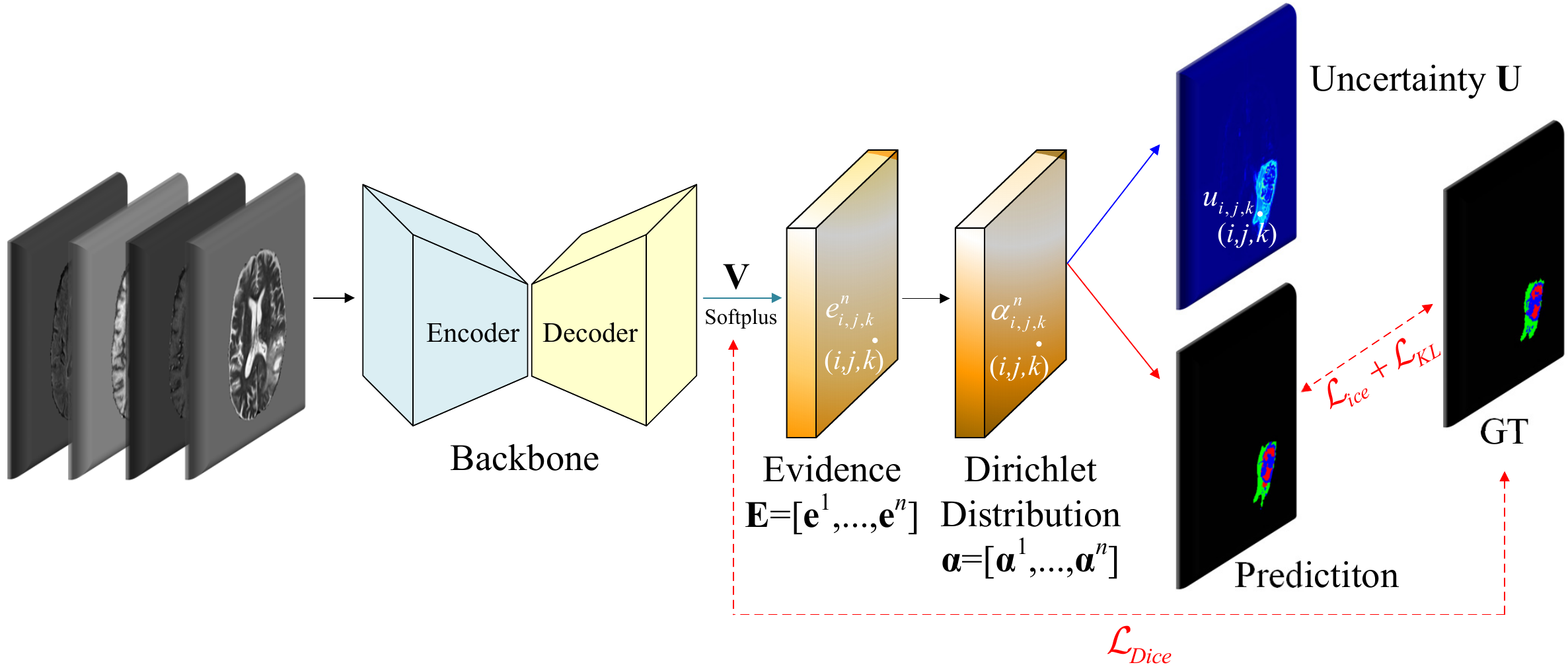}\\
\caption{The framework of trusted brain tumor segmentation.}
\label{F_1}
\end{figure}

One of the generalizations of Bayesian theory for subjective probability is the Dempster-Shafer Evidence Theory (DST) \cite{DST08}. The Dirichlet distribution is formalized as the belief distribution of DST over the discriminative framework in the SL \cite{SL16}. For the medical image segmentation, we define a credible segmentation framework through SL, which derives the probability and the uncertainty of the different class segmentation problems based on the evidence. Specifically, for brain tumor segmentation, SL provides a belief mass and an uncertainty mass for different classes of segmentation results. Accordingly, given voxel-based segmentation results ${\bf{V}}$, its $C + 1$ mass values are all non-negative and their sum is one. This can be defined as follows:
\begin{equation}
\label{E_1}
 \sum\limits_{n = 1}^C {b_{i,j,k}^n} + {u_{i,j,k}} = 1,
\end{equation}
where $b_{_{i,j,k}}^n \ge 0 $ and ${u_{i,j,k}} \ge 0$ denote the probability of the voxel at coordinate $(i,j,k)$ belonging to the $n$-th class and the overall uncertainty value, respectively. In detail, as shown in Fig.~\ref{F_1}, ${\bf{U}}{\rm{ = }}\left\{ {{u_{i,j,k}}, \ \left( {i,j,k} \right) \in \left( {H,W,F} \right)} \right\}$ means the uncertainty for the segmentation results ${\bf{V}}$.  $H$, $W$, and $F$ are the width, height, and number of slices of the input data, respectively. ${{\bf{{b}}}^n}{\rm{ = }}\left\{ {b_{_{i,j,k}}^n, \ \left( {i,j,k} \right) \in \left( {H,W,F} \right)} \right\}$ refers the probability of $n$-th class for the segmentation results ${\bf{V}}$. After that, the evidence ${{\bf{e}}^n}{\rm{ = }}\left\{ {e_{_{i,j,k}}^n, \ \left( {i,j,k} \right) \in \left( {H,W,F} \right)} \right\}$ for the segmentation results ${\bf{V}}$ is acquired by an activation function layer softplus, where ${{e_{i,j,k}^n}} \ge 0 $. Then the SL associates the evidence ${{e_{i,j,k}^n}}$ with the Dirichlet distribution with the parameters ${\alpha _{i,j,k}^n}={{e_{i,j,k}^n}}+1$. Finally, the belief mass and the uncertainty of the $(i,j,k)$-th pixel can be denoted as:
\begin{equation}
\label{E_2}
b_{i,j,k}^n = \frac{{e_{i,j,k}^n}}{S} = \frac{{\alpha _{i,j,k}^n - 1}}{S} \quad{\rm{and }}\quad{u_{i,j,k}} = \frac{C}{S},
\end{equation}
where $S = \sum\limits_{n = 1}^C {\alpha _{i,j,k}^n}  = \sum\limits_{n = 1}^C {\left( {e_{i,j,k}^n + 1} \right)} $ denotes the Dirichlet strength. This describes such a phenomenon that the more evidence of the $n$-th class obtained by the $(i,j,k)$-th pixel, the higher its probability. On the contrary, the greater uncertainty for the $(i,j,k)$-th pixel.

\subsection{Trusted segmentation network} 
The overall architecture of the proposed TBraTS network is shown as Fig. \ref{F_1}. As usual, We adopt the 3D backbone network (V-Net \cite{VNet} or Attention-UNet \cite{AttentionU18}) to obtain the multi-class segmentation results. After that, we construct a trusted  framework with SL \cite{SL16} that induces probabilities and uncertainties for different classes of brain tumor segmentation results. \\
\textbf{Backbone.} Recently, U-Net and its variants were presented to tackle the segmentation problem of medical images. We compared the performances of 3D V-Net and Attention-UNet backbones under our trusted segmentation framework in the experiments. Furthermore, the backbones only performed down-sampling three times to reduce information loss and balance between GPU memory usage and segmentation accuracy. It is worth mentioning that our framework can be freely chosen by the designers with different backbones (such as 3D U-Net and TransUNet~\cite{Transunet21}, etc.).\\  
\textbf{Uncertainty estimation.} For most brain tumor segmentation networks \cite{chen2019disentangle,TransBTS,21kiu}, the predictions are usually obtained by the softmax layer as the last layer. However, it tends to lead the high confidence even for the wrong predictions \cite{ICMLdeterministic20,2020trusted}. Our proposed TBraTS network avoids this problem well in the following way. Firstly, the traditional neural network output ${\bf{V}}$ is followed by an activation function layer to ensure that the network output is non-negative, which is regarded as the evidence voxel ${\bf{E}}$. Second, the SL provides a belief mass function that allows the model to calculate the uncertainty of the segmentation results for different classes. This provides sufficient evidences for clinicians to assist in diagnosing brain tumors. Specifically, as depicted in Fig.~\ref{F_1}, the output result of the backbone network first passes through the layer of softplus, and then calculates the probability and uncertainty of its different categories by Eq.~\ref{E_1} and \ref{E_2}. \\
\textbf{Differences from similar methods.} At last, we analyze the differences of our trusted segmentation network between the traditional brain tumor segmentation methods\cite{TransBTS,21kiu} and the evidential based methods \cite{evidential18,huang2021evidential,huang2021belief}. Compared with the traditional brain tumor segmentation methods\cite{TransBTS,21kiu}, we treat the predictions of the backbone neural network as subjective opinions instead of using a softmax layer to output overconfident predictions.  As a result, our model provides voxel-wise uncertainty estimations and robust segmentation of brain tumor, which is essential for interpretability in disease diagnosis. Compared with the evidential deep learning method \cite{evidential18}, we focus on trusted medical image segmentation and provide uncertainty estimations for 3D voxels. Meanwhile, we develop a general end-to-end joint learning framework for brain tumor segmentation with a flexible backbone network design. Compared with the similar evidential segmentation methods \cite{huang2021evidential,huang2021belief}, we employ the subjective logic theory to explicitly model uncertainty rather than the belief function. Moreover, we verify the robustness of baselines (V-Net and Attention-UNet) with our proposed method under different Gaussian noise for brain tumor segmentation.  

\subsection{Loss function} 
Due to the imbalance of brain tumor, our network is first trained with cross-entropy loss function, which is defined as:
\begin{equation}
\label{E_3}
{{\cal L}_{ce}}=\sum\limits_{n = 1}^C{-{y_m^n}\log\left({p_m^n}\right)},
\end{equation}
where ${y_m^n}$ and ${p_m^n}$ are the label and predicted probability of the $m$-th sample for class n. Then, SL associates the Dirichlet distribution with the belief distribution under the framework of evidence theory for obtaining the probability of different classes and uncertainty of different voxels based on the evidence collected from the backbone. As shown in~\cite{evidential18}, Eq. \ref{E_3} can be further improved as follows:

\begin{equation}
\label{E_4}
\begin{array}{l}
{{\cal L}_{ice}}=\int {\left[ {\sum\limits_{n = 1}^C { - y_m^n\log (p_m^n)} } \right]} \frac{1}{{B({\boldsymbol{\alpha }}_m)}}\prod\limits_{n = 1}^C {p{{_m^n}^{^{{\bf{\alpha }}_m^n - 1}}}d{{\bf{p}}_m}}  \\
\quad\quad {\rm{      }} = \sum\limits_{n = 1}^C {y_m^n\left( {\psi \left( {{\boldsymbol{S}_m}} \right) - \psi \left( {{\boldsymbol{\alpha }}_m^n} \right)} \right)} 
\end{array},
\end{equation}
where $\psi \left(  \cdot  \right)$ denotes the $digamma$ function. ${{\bf{p}}_m}$ is the class assignment probabilities on a simplex, while $B({\boldsymbol{\alpha }}_m)$ is the multinomial beta function for the $m$-th sample concentration parameter ${\boldsymbol{\alpha }}_m$, and ${\boldsymbol{S}_m}$ is the $m$-dimensional unit simplex. More detailed derivations can be referenced in~\cite{evidential18}. To guarantee that incorrect labels will yield less evidence, even shrinking to 0, the KL divergence loss function is introduced by:
\begin{equation}
\label{E_5}
\small
{{\cal L}_{KL}}=\log \left( {\frac{{\Gamma \left( {\sum\nolimits_{n = 1}^C {\widetilde {\boldsymbol{\alpha }}_m^n} } \right)}}{{\Gamma (C)\sum\nolimits_{n = 1}^C {\Gamma \left( {\widetilde {\bf{\alpha }}_m^n} \right)} }}} \right) + \sum\limits_{n = 1}^C {\left( {\widetilde {\boldsymbol{\alpha }}_m^n - 1} \right)} \left[ {\psi \left( {\widetilde {\boldsymbol{\alpha }}_m^n} \right) - \psi \left( {\sum\nolimits_{n = 1}^C {\widetilde {\boldsymbol{\alpha }}_m^n} } \right)} \right],
\end{equation}
where $\Gamma \left(  \cdot  \right)$ is the $gamma$ function. $\widetilde {\boldsymbol{\alpha }}_m^n = y_m^n + \left( {1 - y_m^n} \right) \odot {\boldsymbol{\alpha }}_m^n$ denotes the adjusted parameters of the Dirichlet distribution, which is used to ensure that ground truth class evidence is not mistaken for 0. Furthermore, the Dice score is an important metric for judging the performance of brain tumor segmentation. Therefore, we use a soft Dice loss to optimize the network, which is defined as:
\begin{equation}
\label{E_6}
{{\cal L}_{Dice}}=1 - \frac{{2y_m^np_m^n + \alpha }}{{y_m^n + p_m^n + \beta }},
\end{equation}
So, the overall loss function of our proposed network can be define as follows:
\begin{equation}
\label{E_7}
{\cal L}={{\cal L}_{ice}} + {\lambda _p}{{\cal L}_{KL}} + {\lambda _s}{{\cal L}_{Dice}},
\end{equation}
where ${\lambda _p}$ and ${\lambda _s}$ are the the balance factors, which are set to be 0.2 and 1. 

\section{Experiments}
\textbf{Data \& Implementation Details.} We validate our TBraTS network on the Brain Tumor Segmentation (BraTS) 2019 challenge \cite{BraTSbench,bakas2019}. 335 cases of patients with ground-truth are randomly divided into train dataset, validation dataset and test dataset with 265, 35 and 35 cases, respectively. The four modalities of brain MRI scans with a volume of $240\times240\times155$ are used as inputs for our network. The outputs of our network contain 4 classes, which are background (label 0), necrotic and non-enhancing tumor (label 1), peritumoral edema (label 2) and GD-enhancing tumor (label 4). We combined the three tumor sub-compartment labels to focus on the whole tumor's segmentation results. Our proposed network is implemented in PyTorch and trained on NVIDIA GeForce RTX 2080Ti. We adopt the Adam to optimize the overall parameters with an initial learning rate of 0.002. The poly learning strategy is used by decaying each iteration with a power of 0.9. The maximum of the epoch is set to 200. The data augmentation techniques are similar to \cite{TransBTS}. For the BraTS 2019 dataset, all inputs are uniformly adjusted to $128\times128\times128$ voxels, and the batch size is set to 2. All the following experiments adopted a five-fold cross-validation strategy to prevent performance improvement caused by accidental factors. \\
\textbf{Compared Methods \& Metrics.} Current medical image segmentation methods named U-Net (U) \cite{Unet15}, Attention-UNet (AU) \cite{AttentionU18} and V-Net (V) \cite{VNet} are used for the comparison of the brain tumor segmentation. The following different uncertainty quantification methods are compared with our method. (a) Dropout U-Net (DU) employs the test time dropout as an approximation of a Bayesian neural network \cite{16dropout}. Similar to \cite{MIA2020exploringDropSeg}, DU applied Monte-Carlo dropout ($p$=0.5) on U-Net before pooling or after upsampling. (b) U-Net Ensemble (UE) quantifies the uncertainties by ensembling multiple models \cite{ensemble17}. Although UE shares the same U-Net structure, it is trained with different random initialization on the different subsets (90\%) of the training dataset to enhance variability. (c) Probabilistic U-Net (PU) learns a conditional density model over-segmentation based on a combination of a U-Net with a conditional variational autoencoder \cite{2018probabilisticU}. The following metrics are employed for quantitative evaluation. (a) The Dice score (Dice) is adopted as intuitive evaluation of segmentation accuracy. (b) Normalized entropy (NE), (c) Expected calibration error (ECE) \cite{2019assessing} and (d) Uncertainty-error overlap (UEO) \cite{2019assessing} are used as evaluation of uncertainty estimations.

\begin{figure}[!t]
\centering
\includegraphics[width=0.9\linewidth]{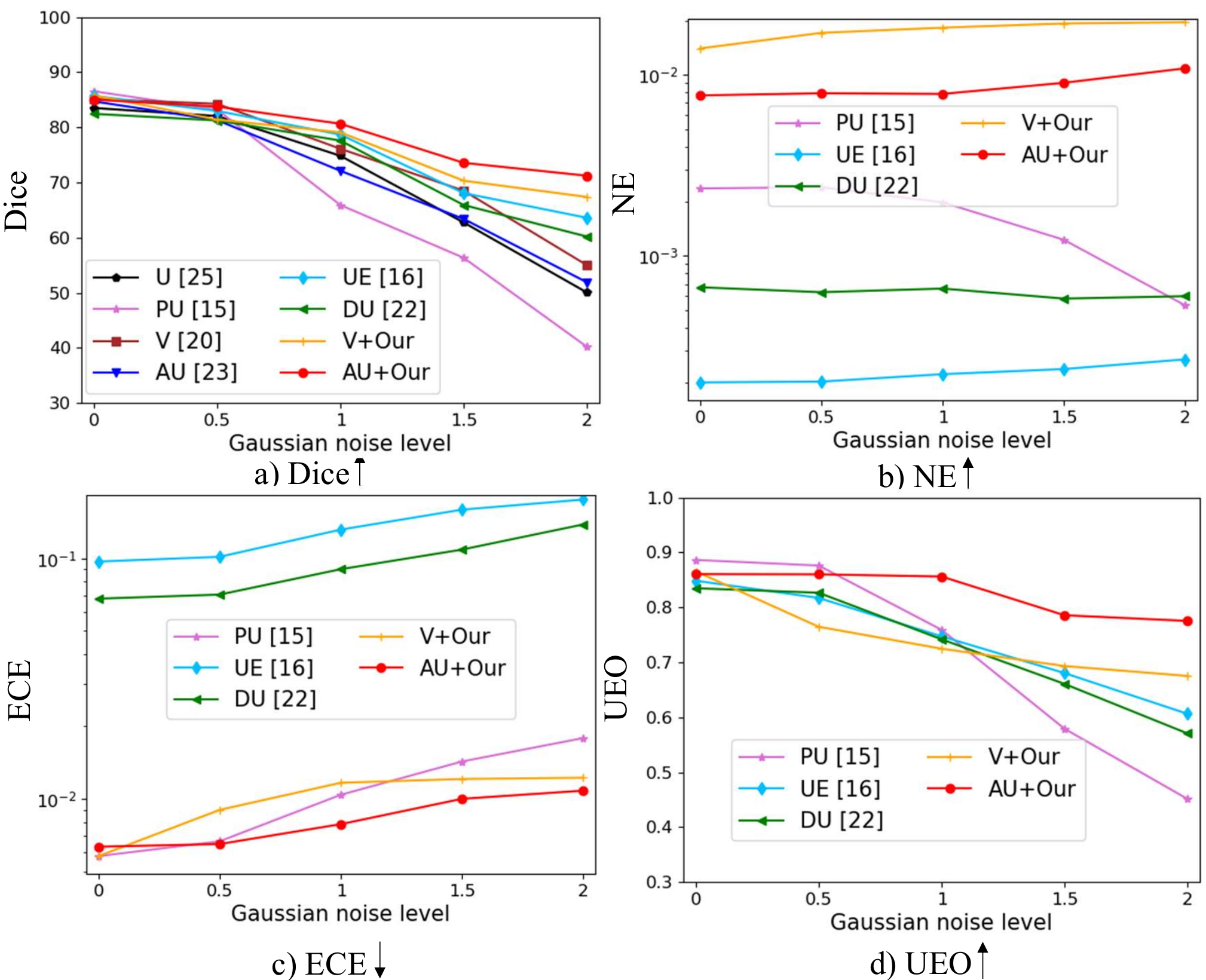}
\caption{The quantitative comparisons with U-Net based methods and uncertainty-based methods on the BraTS2019 dataset under vary noise degradation ($\sigma^2=\left\{ {0,0.5,1,1.5,2} \right\}$).}
\label{F_2}
\end{figure}

\begin{figure}[!t]
\centering
\includegraphics[width=1\linewidth]{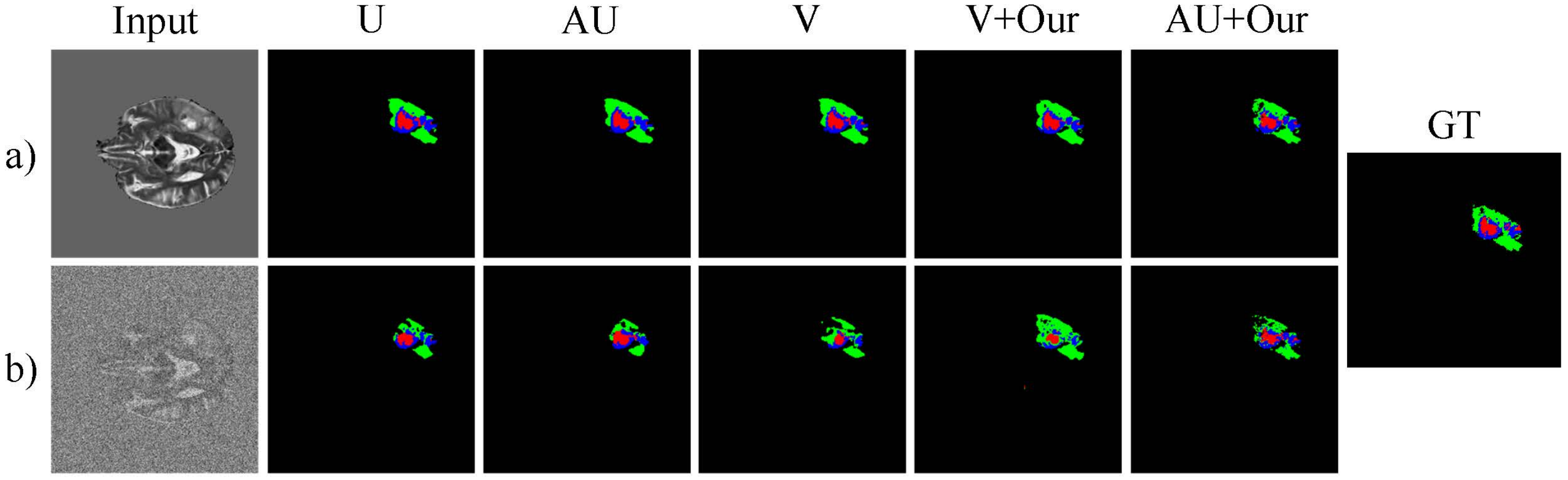}
\caption{The visual comparison of brain tumor segmentation results with different methods. a) Original input (T2 as an example); (b) Noised input under Gaussian noise ($\sigma^2=1.5$).}
\label{F_3}
\end{figure}
\begin{figure}[!t]
\centering
\includegraphics[width=1\linewidth]{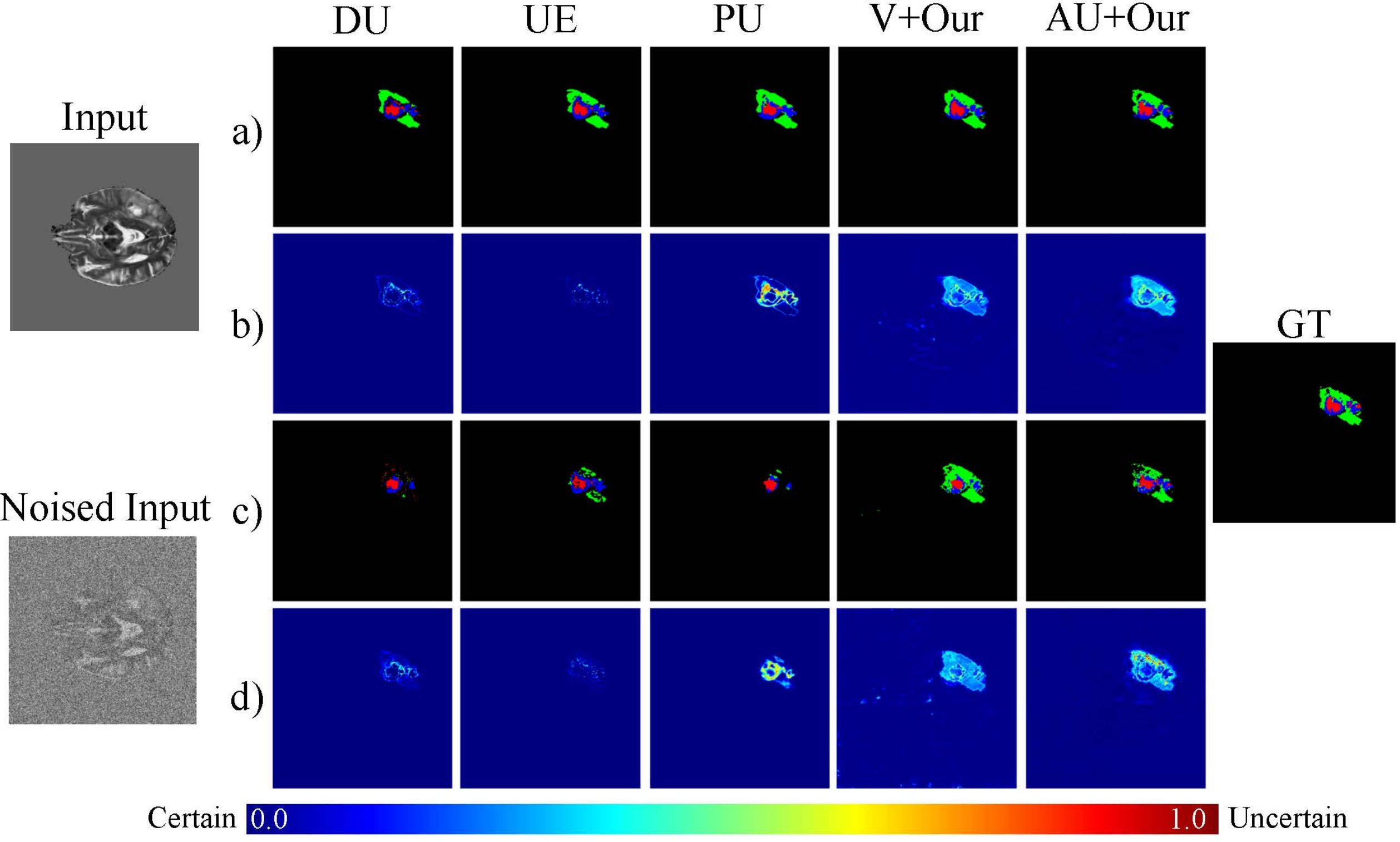}
\caption{The visual comparisons of MRI brain tumor segmentation results with uncertainty-based methods. a) Qualitative results of different methods with the original input (T2 as an example). b) Uncertainty maps for different methods with the original input (T2 as an example). c) Qualitative results of different methods under Gaussian noise with $\sigma^2=1.5$. d) Uncertainty maps for different methods under Gaussian noise with $\sigma^2=1.5$. }
\label{F_4}
\end{figure}

\textbf{Comparison with U-Net based methods.} In Fig. \ref{F_2} (a), we report our algorithm with other U-Net variants on the BraTS 2019 dataset. To verify the robustness of the model, we added Gaussian noise with variance ${\sigma ^2}=1.5$ to the voxels of four modalities. We can observe an interesting fact that when not equipped with our trusted segmentation framework, V-Net and Attention-UNet have shown competitive results with other methods, but their performance drops rapidly with the addition of Gaussian noise. Excitingly, V-Net and Attention-UNet with our framework exhibit more robust performance under increased Gaussian noise. We further show the visual comparison of brain tumor segmentation results under the original input and the high noise input in Fig. \ref{F_3}. It can be seen that V-Net and Attention-UNet with our framework achieve more robust performance than the original backbones. This is attributable to the evidence gathered in the data leading to these opinions.\\
\textbf{Comparison with uncertainty-based methods.} To further quantify the reliability of uncertainty estimation, we compare our model with different uncertainty-based methods, using the elegant uncertainty evaluation metrics of NE, ECE and UEO. As depict in Fig. \ref{F_2} (b)-(d), the performance of all uncertainty-based methods decay gradually under increasing levels of Gaussian noise. Fortunately, our method decays more slowly with the benefit of the reliable and robust evidences captured by our trusted segmentation framework. The test running time of the uncertainty-based method on one sample is 0.015 $mins$ (AU+Our), 0.084 $mins$ (V+Our), 0.256 $mins$ (PU), 1.765 $mins$ (DU) and 3.258 $mins$ (UE). It can be concluded that the running time of our framework is lower than other methods. This is due to the fact that PU and DU will sample at test time to obtain uncertainty estimations, while the UE obtains uncertainty estimations by ensembling multiple models. Moreover, to more intuitively demonstrate the reliability of uncertainty estimations, we show a visual comparison of brain tumor segmentation results from various methods. As shown in the fourth and fifth columns of Fig. \ref{F_4}, the V-Net and Attention-UNet equipped with our framework obtain more accurate and robust uncertainty estimations, even under strong Gaussian noise. The reason being the following two points, we did not use softmax for output which would lead to over-confidence; we employ a subjective logical framework to gather more favorable and robust evidences from the input.

\section{Conclusion}
In this paper we presented an end-to-end trusted segmentation model, TBraTS, for reliably and robustly segmenting brain tumor with uncertainty estimation. We focus on producing voxel-wise uncertainty for brain tumor segmentation, which is essential to provide confidence measures for disease diagnosis. The theory of subjective logic is adopted to model the predictions of the backbone neural network without any computational costs and network changes. Furthermore, Our model learns predicted behavior from the perspective of evidence inference, through the connection between uncertainty quantification and belief mass of the subjective logic. Extensive experiments demonstrated that TBraTS is competitive with previous approaches on the BraTS 2019 dataset.

\small{\textbf{Acknowledgements:} This work was supported by A*STAR Advanced Manufacturing and Engineering (AME) Programmatic Fund (A20H4b0141); Miaozi Project in Science and Technology Innovation Program of Sichuan Province (2021001).}

%
%
%
\bibliography{samplepaper}
\bibliographystyle{splncs04}
\end{document}